\documentclass[12pt,a4paper]{article}
\usepackage{bbm}
\usepackage{amsmath}
\usepackage{amssymb}
\usepackage{graphicx}
\usepackage{psfrag}
\begin{document}

\begin{flushright}
BI-TP 00/18\\
\end{flushright}

\begin{center}  
\large{\bf Large $N$ limit of the IKKT matrix model}
\end{center}

\begin{center}  
P. Bialas$^{1,2}$, Z. Burda$^{1,3}$, B. Petersson$^{1}$, J.~Tabaczek$^{1}$
\end{center}

\centerline{$^{1}$Fakult\"at f\"ur Physik, Universit\"at Bielefeld} 
\centerline{P.O.Box 100131, D-33501 Bielefeld, Germany}
\vspace{0.3cm}
\centerline{$^{2}$Inst. of Comp. Science, Jagellonian University} 
\centerline{33-072 Krakow, Poland}
\vspace{0.3cm}
\centerline{$^{3}$Inst. of Physics, Jagellonian University} 
\centerline{33-059 Krakow, Poland}
\vspace{0.3cm}

\begin{abstract} \normalsize \noindent
Using the dynamical triangulation approach we perform a numerical study
of a supersymmetric random surface model that corresponds to the large
$N$ limit of the four--dimensional version of the IKKT matrix model.
We show that the addition of fermionic degrees of freedom suppresses the
spiky world-sheet configurations that are responsible for the pathological
behaviour of the purely bosonic model. We observe that the distribution of
the gyration radius has a power--like tail $p(R) \sim R^{-2.4}$.
We check numerically that when the number 
of fermionic degrees of freedom is not susy--balanced,
$p(R)$ grows with $R$ and the model is not well--defined.
Numerical sampling of the configurations in the tail of the
distribution shows that the bosonic degrees of freedom collapse to a
one--dimensional tube with small transverse fluctuations. Assuming that the
vertex positions can fluctuate independently within the tube, we give a
theoretical argument which essentially explains the behaviour
of $p(R)$ in the different cases, in particular predicting
$p(R) \sim R^{-3}$ in the supersymmetric case.
Extending the argument to six and ten dimensions, we predict
$p(R) \sim R^{-7}$ and $p(R) \sim R^{-15}$, respectively. 
\end{abstract}

\section*{Introduction}

During the last decades we have witnessed a rapid development
of methods of geometrical quantization. It was triggered by
the discovery of the integration measure over
two-dimensional geometries which was applied
in calculations of quantum amplitudes in string theory \cite{p}. 
In practice, it turned out that the geometrical
quantization of strings works only for either non-physical dimensions
$d \le 1$ \cite{kpz}, or in the critical dimension $d = 10$ 
where the world-sheet degrees of freedom of the
string decouple from the theory \cite{p, gswp}.

The case of non-critical strings with $d \le 1$ was also
independently formulated in terms of a matrix model which 
is equivalent to the dynamical triangulations approach 
\cite{dtle1}. This method allows for calculating different
topological contributions in the double scaling limit \cite{ns}.
However, both the Polyakov conformal field approach and the
matrix model calculations break down at $d = 1$. The problems
that turn up at this point, also known as the `$d = 1$ barrier',
are attributed to instabilities of the conformal mode.

Whether it is possible to stabilize the conformal mode for
surfaces embedded in physical dimensions is not yet clear. Some
attempts to solve this problem have been made within the
conformal field formalism \cite{p2}. Another idea is to extend
dynamical triangulations to the sypersymmetric case, and to attack 
the problem directly in physical dimensions. 
By introducing supersymmetry one hopes to avoid the 
instabilities of bosonic surfaces that manifest
as the degeneration of the world-sheet geometry into branched
polymers \cite{bp}. 

Disrectization of supersymmetric surfaces is known to be 
a difficult problem. A discretization of the world-sheet 
supersymmetry, for example, inherits all the difficulties 
which appear arleady on regular lattices
\cite{gp} and on top of this it introduces new problems related to
the fact that the symmetry must be local \cite{bjk}.
Supersymmetry is broken explicitly by the lattice, and it is not
clear how to ascertain whether it gets restored in the continuum
limit. One encounters similar problems for the Green-Schwarz type
of surfaces where one has to preserve the $\kappa$-symmetry, which
is a local world-sheet symmetry \cite{msav}. One also attempted to 
introduce supersymmetry into the surfaces models indirectly
by considering models with effective actions with extrinsic curvature 
terms obtained by integrating out fermions from the supersymetric
theory \cite{w}. Such discretized surfaces have been extensively 
studied (see e.g. \cite{ec}), but it is not clear what is nature
of the critical point observed in the discretized theory and whether
it can be directly related to the original supersymetric theory
of continuum surfaces. 

In this paper we study a model of supersymmetric surfaces that constitutes
a sort of classical limit of the IKKT matrix model \cite{ikkt} obtained by
sending the matrix size $N$ to infinity \cite{ikkt,bars}. 
The attractive feature of the surface model is that the supersymmetry
is introduced in such a way that in principle the model can be
straightforwardly discretized. By re-writing the IIB string action in
the Schild gauge \cite{ikkt}, one avoids the problems related to the
proper treatment of the local $\kappa$-symmetry in the discretization.
The action that results from this procedure has
no redundant degrees of freedom left, and no local symmetry.

We investigate this model by means of the dynamical
triangulation approach. Some preliminary studies in this direction
were already done in \cite{oy}. We perform Monte-Carlo simulations
of surfaces embedded in a 4--dimensional target space. We find a
theoretical picture that explains the numerical results for $d=4$,
and allows us to make some predictions for the behaviour of the
ten--dimensional version of the model.  The results presented here
give an insight into typical geometrical features of surfaces that
play an important role in the ensemble generated by the IKKT matrix model
\cite{ikkt}. 

The paper is organized as follows. First we shortly recall the model
and describe the discretization scheme. Then we discuss singularities
of the purely bosonic model and show that they are removed by adding
fermions.  For the model with fermions we estimate the large $L$
behaviour of the partition function by assuming that it is dominated by
tube--like configurations. We present the results of the $MC$ simulations
to show that tubes do indeed dominate in the ensemble. We conclude the
paper with a short discussion, concerning in particular the relation
of the results to the matrix model \cite{ikkt,mm,ks}.

\section*{The model}

The action for IIB strings can be cast into the following form
\cite{ikkt}~:
\begin{eqnarray}
S (g, X, \bar{\Psi}, \Psi) & = & \int d^2 \xi \sqrt{g} \
\left( \frac{1}{4} \{ X^\mu, X^\nu \}^2          
- \frac{i}{2}
\bar{\Psi} \Gamma_\mu \{ X^\mu, \Psi \} + \lambda \right)
\label{caction}
\end{eqnarray}
Here $g_{ab}$ is the metric tensor on the world-sheet; $X^\mu$ are
bosonic co-ordinates in the $d$-dimensional target space; $\bar{\Psi}$
and $\Psi$ are $k$-dimensional spinors; and $\Gamma_\mu$ are $k \times
k$ Dirac matrices. For $d = 10$, the dimensionality of the spinor
representation is $k = 32$, and $\bar{\Psi}, \Psi$ are Majorana-Weyl
spinors.  For $d = 4$, we have $k = 4$, and $\bar{\Psi}, \Psi$ are
Weyl spinors. The Poisson brackets in \eqref{caction} are defined as
\begin{equation}
\{ X, Y \} = \frac{1}{\sqrt{g}} \ \epsilon^{ab} \,
\partial_a X \, \partial_b Y \, .
\end{equation}
Based on this action the authors of \cite{ikkt} introduced a matrix
model which was then interpreted as a constructive definition of
superstring theory. The matrix model is obtained from \eqref{caction}
by substituting fields $X^\mu(\xi_1,\xi_2)\rightarrow A^\mu_{ab}$ and 
$\Psi(\xi_1,\xi_2)\rightarrow \Phi_{ab}$ by $N\times N$ traceless
Hermitean matrices, and the Poisson brackets by commutators
$\{,\} \rightarrow -i [,]$. The resulting theory
\begin{equation}
{\cal S}
= - \mbox{Tr} \left( \frac{1}{4} [ A^\mu, A^\nu ]^2
+ \frac{1}{2} \bar{\Phi} \Gamma_\mu [ A^\mu, \Phi ] -
\lambda \mathbbm{1} \right)
\label{maction}
\end{equation}
is quantized by the Feynman integral
\begin{equation}
Z = \int \rm{d} A \ \rm{d} \bar{\Phi} \ \rm{d} \Phi  \ e^{-{\cal S}}
\end{equation}
with the standard integration measure for traceless Hermitean
matrices. This
procedure defines a kind of third quantization of string theory,
which is supposed to reproduce in a double scaling limit the sum
over all topological and geometrical excitations of the world-sheet
of strings \cite{ikkt,bars}. The original model \cite{ikkt} was
defined in ten dimensions.

Here we will discuss the semiclassical limit of $N \rightarrow \infty$,
in which the model reduces to the string theory on a fixed topology
\cite{bars} with the partition function
\begin{equation}
Z = \int D\sqrt{g} D X D \bar{\Psi} D \Psi \,
e^{- S (g, X, \bar{\Psi}, \Psi)}
\end{equation}
where the action is given by \eqref{caction}. We will concentrate for now
on the case $d = 4$, where we can most easily perform numerical simulations.

We use the dynamical triangulation approach to regularize the integration
measure $D\sqrt{g}$. The partition function is then given by
\begin{equation}
Z = \sum_{N} e^{-\lambda N} z_N
\end{equation}
where
\begin{equation}
z_N = \sum_{T \in {\cal T}_N} \frac{1}{C_T}
\int \sideset{}{'}\prod_i^n d^4 X_i
\sideset{}{'}\prod_i^n d^2 \bar{\Psi}_i d^2 \Psi_i
\ e^{- S_T (X_i, \bar{\Psi}_i, \Psi_i)} \quad .
\label{zz}
\end{equation}
The sum runs over equilateral, oriented triangulations $T$
with $N$ triangles and $n = N / 2 + 2$ vertices. $C_T$ is the symmetry
factor of a given $T$, and $S_T$ is the discretized action on this
particular triangulation. Primes on the products indicate that the
zero modes were removed (see the following discussion). The fields
$X_i$ and $\Psi_i$ are now located on the vertices of $T$. The action
has one parameter, namely the cosmological constant $\lambda$. The
continuum limit is taken by adjusting $\lambda$ in the grand-canonical
partition function $Z(\lambda)$ to its critical value and sending the
lattice spacing $a$ to zero. Alternatively, if one prefers to work
with the ensemble of canonical partition functions $z_N$, one can take
the continuum limit by simultaneously sending $N$ to infinity and $a$
to zero while holding the physical area of the worldsheet $A \sim N
a^2$ fixed. This latter approach is the one we will choose here.
 
\begin{figure}
\begin{center}
\includegraphics[height=5cm]{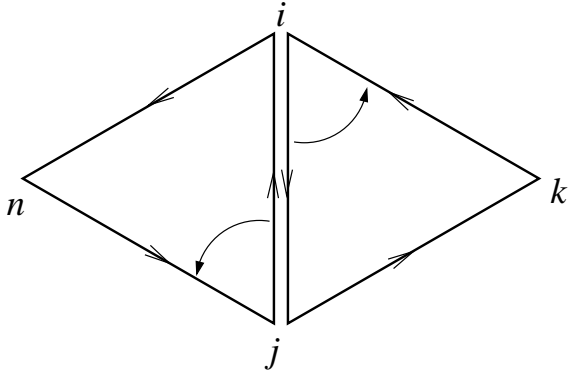}
\end{center}
\caption{\label{omega} Two neighbouring, oriented triangles
$\langle ijk \rangle$ and $\langle jin \rangle$, connected by the
link $\langle ij \rangle$. The `outer' vertices $k$ and $n$
can be accessed by the operation $\omega$ as defined in the text,
with $\omega (ij) = k$ and $\omega (ji) = n$.}
\end{figure}

The discretized action consists of two terms, a purely bosonic part
$S_B$ and a fermionic part $S_F$. Following the suggestions in
\cite{oy}, these are~:
\begin{eqnarray}
S_B & = & {\cal B} \sum_{\langle ijk \rangle} \left\{
- \left( (X_{ij}^2)^2 + (X_{jk}^2)^2 + (X_{ki}^2)^2 \right) \right.
\nonumber \\
& & \left. \ \ \ \ \ \ \ \ \ + 2 \left( X_{ij}^2 X_{jk}^2
+ X_{jk}^2 X_{ki}^2 + X_{ki}^2 X_{ij}^2 \right) \right\} \label{bact} \\
S_F & = & \frac{i}{12} \sum_{\langle ij \rangle}
\bar{\Psi}_i \Gamma_\mu \Psi_j (X_{\omega(ij)}^\mu - X_{\omega(ji)}^\mu) \ .
\end{eqnarray}
In contrast to the action presented in \cite{oy},
here $\bar{\Psi},\Psi$ are Weyl fermions.
The first sum runs over all triangles of the triangulation, the
second over all links. The constant in \eqref{bact} is set to
${\cal B} = (8\sqrt{3} a^2)^{-1}$. The operation $\omega$, when
acting on an oriented link $\langle ij \rangle$, gives that neighbour
of $i$ that comes after $j$ when going counterclockwise around $i$.
In other words, if we denote two neighbouring oriented triangles
by $\langle ijk \rangle$ and $\langle jin \rangle$, then
$\omega (ij) = k$ and $\omega (ji) = n$ (see figure \ref{omega}).

Two things should be noted about this action. Firstly, the bispinors
$\bar{\Psi},\Psi$ in the target space have a handedness, {\em i.e.}
$\Psi = \frac{1}{2} (\mathbbm{1} + \Gamma_5) \Psi$ and
$\bar{\Psi} =  \bar{\Psi} \frac{1}{2} (\mathbbm{1} - \Gamma_5)$.
Using the chiral representation of the Dirac matrices in the Euclidean
sector,
\begin{equation}
\vec{\Gamma} = \left( \begin{array}{ll}
0 & \vec{\sigma} \\ \vec{\sigma} & 0 \end{array} \right) \ , \quad
\Gamma_0 = \left( \begin{array}{ll}
0 & - i \sigma_0 \\ i \sigma_0 & 0 \end{array} \right) \ , \quad
\Gamma_5 = \left( \begin{array}{ll}
\sigma_0 & 0 \\ 0  & -\sigma_0 \end{array} \right)
\end{equation}
where $\vec{\sigma}$ are the Pauli matrices and $\sigma_0$ is 
the $2\times 2$ unity matrix, we can parametrize the chiral bispinors
$\bar{\Psi}$ and $\Psi$ in terms of two-component spinors
$\psi, \bar{\psi}$~:
\begin{equation}
\bar{\Psi} = (0 , \bar{\psi}) \ , \quad
\Psi = \left(\begin{array}{c} \psi \\ 0 \end{array}\right) \ .
\end{equation}
Thus, we can re-write the fermionic part of the action as
\begin{equation}
S_F (X, \bar{\psi}, \psi) = \sum_{\langle ij \rangle}
\bar{\psi}_i \left[ i \vec{f}_{ij} \vec{\sigma}
- f^0_{ij} \sigma_0 \right] \psi_j 
\label{sf}
\end{equation}
where $f^\mu_{ij} = \frac{1}{12} \left( X^\mu_{\omega(ij)}
- X^\mu_{\omega(ji)} \right)$ are real numbers that are antisymmetric
in $i$ and $j$, $f^\mu_{ij} = - f^\mu_{ji}$.

Secondly, the action has a zero mode related to the translational
invariance in the bosonic sector, $X^\mu_i \rightarrow X^\mu_i + \delta$.
As a remnant of the original supersymmetry, there is also a similar
symmetry in the fermionic sector~: $\bar{\psi}_i \rightarrow \bar{\psi}_i
+ \bar{\epsilon}$, $\psi_i \rightarrow \psi_i + \epsilon$ \cite{oy}. This
might be not evident at first sight, but the change in the action is
indeed zero,
\begin{eqnarray}
\delta S_F & = &
\sum_{\langle ij \rangle} \bar{\epsilon}
\left( i \vec{f}_{ij} \vec{\sigma} - f^0_{ij} \sigma_0 \right) \psi_j 
\ + \ \sum_{\langle ij \rangle} \bar{\psi}_i
\left( i \vec{f}_{ij} \vec{\sigma} - f^0_{ij} \sigma_0 \right) \epsilon
+ \nonumber \\
& & \ + \ \sum_{\langle ij \rangle} \bar{\epsilon}
\left( i \vec{f}_{ij} \vec{\sigma} - f^0_{ij} \sigma_0 \right) \epsilon = 0
\end{eqnarray}
because for each vertex $i$ the sum of the $f_{ij}$ over all
its neighbours $j$ is zero~: $\sum_j f^\mu_{ij} = 0$.

These zero modes have to be divided out in the measure. In practice,
this can be done by inserting a delta function $\delta^d (X_n)$ and
an additional product $\bar{\psi}_n \psi_n$ -- which acts like a delta
function for the Grassmann variables $\bar{\psi}_n, \psi_n$ -- for one
vertex.

Taking all this into account, the canonical partition function 
for the four dimensional model now reads
\begin{equation}
z_N = \sum_{T \in {\cal T}_N} \frac{1}{C_T} \int \prod_i^{n - 1} d^4 X_i
\prod_i^{n} d^2 \bar{\psi}_i d^2 \psi_i \cdot
\bar{\psi}_n \psi_n \cdot 
e^{- S_B (X)} e^{- \sum_{\langle ij \rangle} \bar{\psi}_i a_{ij} \psi_j}
\, ,
\label{part}
\end{equation}
where
\begin{equation}
a_{ij} = i \vec{f_{ij}} \vec{\sigma} - f^0_{ij} \sigma_0 \, .
\label{defa}
\end{equation}
Integration over the fermions results in the appearance of a factor $D(X)
= \det a'_{ij}$, where $a'_{ij}$ is a $2 (n - 1) \times 2 (n - 1)$
matrix that is obtained from $a_{ij}$ by crossing out the two rows and
columns that correspond to the vertex $n$. The determinant $D(X)$ is a
non-negative function of bosonic coordinates $X$, a feature that is
essential for the MC simulations. It follows from the structure of 
the matrix $a'_{ij}$, which has pairs of complex
conjugated eigenvalues $\lambda, \lambda^*$. (More specifically, if
$\eta_i$ is an eigenvector to the eigenvalue $\lambda$, then $\sigma_2
\eta^*_i$ is an eigenvector to the eigenvalue $\lambda^*$.) 

The final form of the partition function is now
\begin{equation}
z_N(\gamma) = 
\sum_{T \in {\cal T}_N} \frac{1}{C_T} \int \prod_{i}^{n - 1} d^4 X_i \,
D(X)^{\gamma} \, e^{- S_B (X)}
\label{pf}
\end{equation}
where for convenience we introduced an additional parameter $\gamma$.
For $\gamma = 0$ we have the purely bosonic model; $\gamma = 2$ is the case
considered in \cite{oy}; and $\gamma = 1$ corresponds to our partition
function \eqref{part}. As we will see, the model with $\gamma = 1$,
for which the number of bosonic and fermionic degrees of freedom is
susy--balanced, is the only interesting one, because only in this case
is the partition function \eqref{pf} well-defined.

\section*{Singularities of the bosonic partition function}

The behaviour of the partition function for the purely bosonic
model $z_N (0)$ has already been discussed in the context of
the quantization of the Nambu-Goto string \cite{adf}, where it
was shown that this partition function is in general not
integrable. Namely, one can explicitly construct triangulations
for which the integral over bosonic fields is divergent. 

\begin{figure}
\begin{center}
\psfrag{r}{$\sim r$}
\includegraphics[width=5cm]{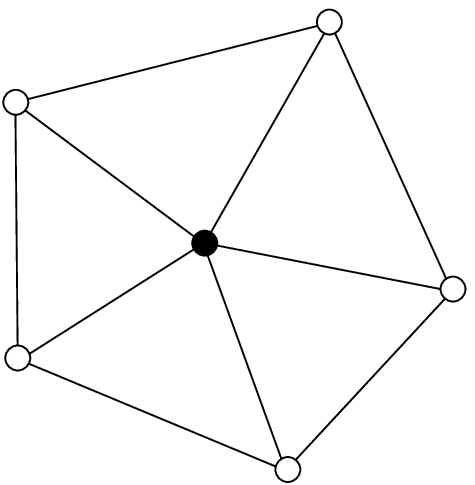}
\kern4mm\includegraphics[width=5cm]{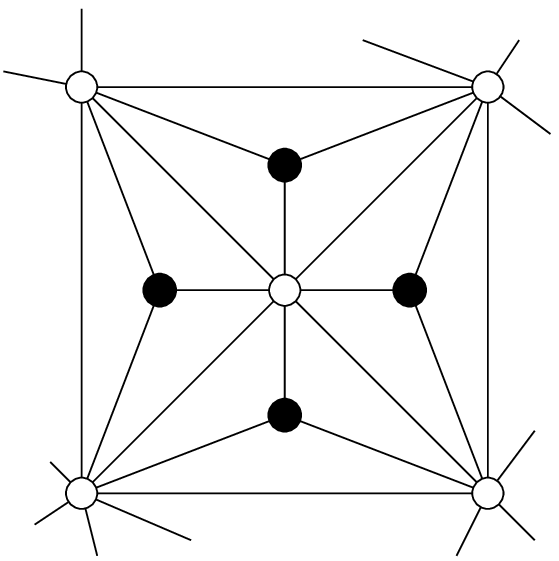}
\end{center}
\caption{\label{spikes} Origin of spikes~: (Left) When the white points
approach each other to a distance $\sim r$, then the black point can 
freely move in the embedding space inside a sphere of radius $\sim 1/r$.
(Right) Example of the non--integrable configuration : the entropy
factor from four spikes (black points) overcomes the damping effect of
the entropy that comes from the white points which are forced to move
close to each other.}
\end{figure}

The source of this divergence can be intuitively explained as
follows. The bosonic part of the action is the sum of all triangle
areas squared, with the areas measured in the target space. Thus, as
long as the areas of all triangles stay constant, we can change their
shape without changing the action. As an example, let us pick an
arbitrary vertex $m$ on the triangulation. It has $q$
neighbouring vertices that are linked to each other, forming a loop
around $m$ (see figure \ref{spikes} (left)).
Each link in this loop is the base of one of $m$'s neighbouring
triangles. Now let us shrink these bases to a small length $r$.
(Obviously, this step {\em will} change the action, but this is a {\em
finite} and {\em local} change that is bounded independently of $r$.)
Next, we can move $m$ outward to a distance of about $\sim 1/r \equiv L$,
thus creating a long `spike'. In this way, we can have an arbitrarily long
spike that does not visibly affect the action if we just
simultaneously make the bases small enough.

The phase space volume for such a configuration is a factor
$L^d$ coming from the point $m$ times a factor 
$r^{(q-1)d} = L^{-(q-1)d}$
coming from constraining the $q$ neighbours of $m$ 
to a sphere of radius $r$ around one of them. 
Thus, a spike of length $L$ is suppressed, but only 
{\em power--like}, {\em i.e.} the spikes are
not cut off at any particular scale and arbitrarily long fluctuations
can occur. When instead of a single--vertex spike one considers
configurations with multiple spikes, such as the one in figure
\ref{spikes} (right), one can check that the spike length distribution
may have a logarithmic or stronger divergence.

\section*{Singularities of the susy partition function}

The question that now arises is, what can be expected to change
in this picture if we add the fermionic part to the action and
consider the complete partition function $z_N (1)$? Intuitively,
it seems that fermions should introduce a repulsive core to the
effective potential between neighbouring vertices on the triangulation,
and that this repulsion should prevent vertices from occupying the
same position in the target space, which in turn should keep the
loop around each vertex from shrinking beyond a certain minimal
length. In this case, to create a spike would cost energy, because
now the area of a triangle would have to grow to allow it to become
elongated. Therefore, one can hope that the introduction of fermions
might suppress spikes, along with the divergences they cause.

Introducing the fermions is equivalent to replacing the bosonic
action $S_B$ with an effective action
\begin{equation}
S_{B, eff} (X) = S_B (X) - \log D (X)
\label{eff}
\end{equation}
where $D (X) = \det a'_{ij}$ is the determinant of the matrix defined in
\eqref{defa}. $D (X)$ vanishes if for any vertex all its neighbouring
vertices meet in a single point, because then the two rows of the matrix
that correspond to this vertex become zero. Similarly, if all neighbouring
vertices do not exactly meet in one point but move very close to each
other, then all entries in these two rows become very small, and the
determinant as a whole should likewise become small. In other words,
we expect the determinant to discourage small loops and hence also spikes.
It is natural to expect that short loop lengths will be suppressed
by a power-like fall-off.

To test this, we used standard Monte Carlo methods to simulate the
model with fermions.  The triangulations, weighted by the effective
action \eqref{eff}, were produced by a Metropolis algorithm for
geometrical flips and updates of the fields $X^\mu$. In four
dimensions the effective
action is well-defined since the determinant is positive definite as
discussed before. The bulk of the computer time is spent on
calculating the determinant $D(X)$. This allowed us to produce
reasonably large statistics only for fairly small systems (up to
16 vertices).

\begin{figure}
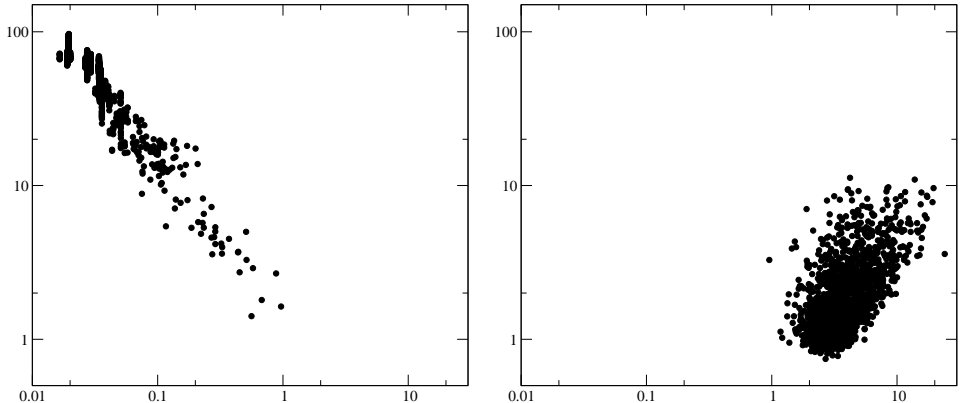

\begin{center}
\psfrag{xl}{}
\psfrag{yl}{}
\includegraphics[clip,width=6.5cm]{loop_spikes_0.eps}\includegraphics[clip,width=6.5cm]{loop_spikes_1.eps}
\caption{\label{loop_spikes} 
The gyration radius versus the shortest loop around a vertex 
on the triangulation, for the model without fermions (left)
and with fermions (right). The number of vertices is $n = 16$.}
\end{center}
\end{figure}

The simplest quantity to measure is the average action.
Using the fact that the bosonic action and the fermionic
determinant are uniform polynomials of the bosonic co-ordinates,
one can show by a simple scaling argument that the
average bosonic action for $d = 4$ is
$\langle S_B \rangle = (n - 1) (1 + \gamma / 2)$.
We used this formula as a rough test of the program.

\begin{figure}
\begin{center}
\includegraphics[clip,width=6.5cm]{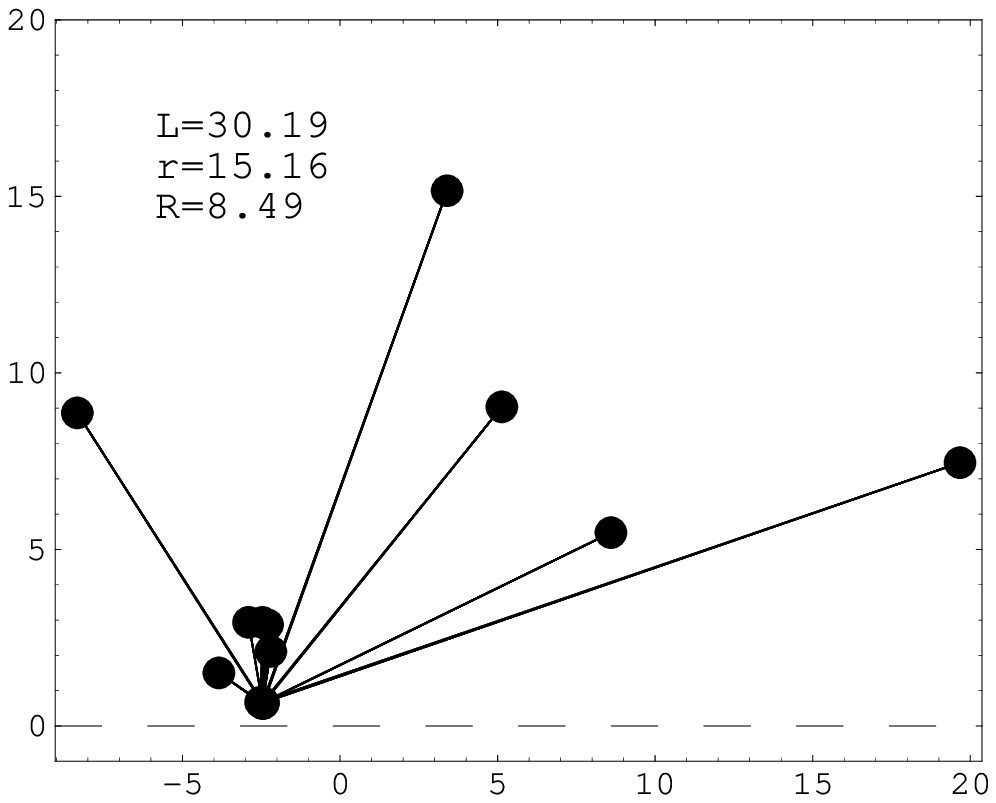}\includegraphics[clip,width=6.5cm]{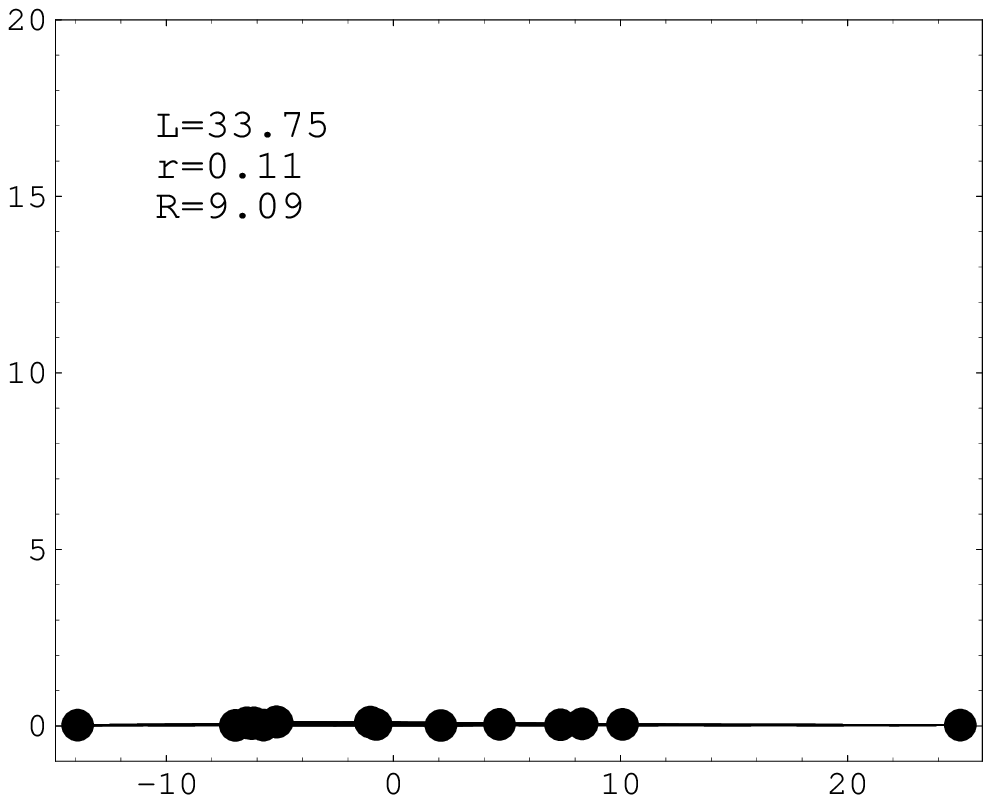}
\caption{\label{needle} Snapshots of a susy triangulation from the tail
of the gyration radius distribution (right), and a bosonic configuration
of comparable size (left). The $x$-component in the plot 
corresponds to the projection of the vertex position onto
the longest principal axis in the system, while the $y$-component 
describes the distance from the axis. $L$ denotes the length of the tube,
$r$ its radius and $R$ the gyration radius} 
\end{center}
\end{figure}

We also measured the gyration radius
\begin{equation}
R = \sqrt{\frac{1}{n} \sum_i^n (X_i - X_{CM })^2} 
\label{R2}
\end{equation}
and the length of the smallest loop around a vertex that appears on
each configuration. (`Length', here, means the sum of the squared
lengths of all links in the loop.)  Figure \ref{loop_spikes} (left) shows
that in the purely bosonic model there is a strong correlation between
the appearance of short loops on the triangulation and a large
gyration radius, which supports the picture discussed in the previous
section. Note that all results for the bosonic model should be taken
with a generous helping of salt, since as noted above the partition
function is actually not defined. These results are presented here
merely to illustrate the fact that the source of the singularities can
really be found in the appearance of local spikes (see figure
\ref{needle} (left)). For the fermionic case, small loops are clearly
suppressed, and the only correlation between loop lengths and gyration
radius is a trivial one, namely that if the shortest link on a
triangulation has a length $r$, the gyration radius itself cannot be
smaller than $r$, either. We conclude that fermions do in fact
suppress spikes. For the smallest possible spherical
triangulation, the tetrahedron, one can actually show analytically
that the partition function for the fermionic model exists in four
dimensions, whereas for the bosonic model it does not \cite{oy}.

However, both the analytical calculations for the tetrahedron and numerical
simulations of larger surfaces indicate that even in the fermionic case
the gyration radius distribution has a power--like tail (see the appendix
and figure~\ref{fig:hist} (left)).
What is the source of the singular behaviour in this case, when we
know that we do not have local spikes? It turns out that there is
another type of configuration that can contribute to the tail of the
probability distribution $p(R)$ for large $R$. These things look like
{\em needles}; the four-dimensional degrees of freedom of the co-ordinates
in the target space collapse to a one-dimensional, elongated narrow
tube.
To indentify these configurations we used the principal
component analysis (PCA).  For each analysed configuration we
calculated the correlation matrix $Q_{\mu\nu}=\frac{1}{n}\sum_i{X^\mu
X^\nu}$ and its eigenvalues.  The eigenvector corresponding
to the largest eigenvalue defined the principial axis. For each point on
the configuration $X_i$ we then calculated its projection on the pricipial
axis, $X^L_i$, and its distance from it, $d_i$. We defined the length $L$
and radius $r$ of the tube by 
\begin{equation}
L=\max_i X^L_i-\min_i X^L_i \quad\text{and}\quad r= \max_i d_i
\end{equation}
The ratio of length to width can serve as an estimator of the type
of the configuration. For the bosonic string, this ratio was $2.26(8)$,
and for the fermionic one, $212(20)$.  Two typical results of this analysis
are depicted in figure \ref{needle} where we show, for comparison,
snapshots of triangulations for the bosonic (left) and susy (right)
models for $n=16$, projected onto the principal axis.

How can the appearance of needles be explained? The bosonic part of
the action \eqref{bact} is a sum of contributions from separate triangles.
The contribution from a single triangle is minimal when its three vertices
lie on a line, in which case the square of the triangle area is zero.
One type of zero-action triangles are the spikes that were discussed
in the previous section. On a spike, two vertices of the triangle lie
very close to -- or, in the limiting case, on top of -- each other, so
that together with the third vertex they will automatically form a line.
Another possible type of zero-action triangles are those that have three
distinct (and distant) points that lie on, or close to, a
single line. From the point of view of just one triangle, this type of
zero-action configuration is more likely to occur than a spiky one, for
which one has to move two vertices into exactly the same point; but when
one looks at the triangulation as a whole, one sees that the probability
for this to happen is actually much smaller, because having three long links
means that either there will be three neighbouring triangles with large
areas (which is suppressed by the action), or there will be three more
triangles lying on exactly the same line. The same argument can then be
applied to the neighbours of {\em these} triangles as well, and on and on
until one arrives at a configuration where all triangles lie on a single
line, {\em i.e.} a configuration that is just one-dimensional. Contrary
to spikes, which are defects that involve only a few triangles and that
are basically independent of the rest of the triangulation, a needle
requires a global arrangement of vertices that minimizes the action.
Such a global arrangement is entropically disfavoured and therefore does 
not occur in the bosonic model. Here, however, the fermionic determinant
provides an additional factor that favours configurations with vertices
far removed from each other. As it turns out, this term is sufficient to
exceed the damping effect coming from the entropy.

To see this, we will try to estimate the contribution of the needle--like
configurations to the partition function. Let us assume that we fix one
point on the surface (thus getting rid of the zero mode), and also set
the length of one link emerging from this point to $L$. In the
following, we will refer to the direction along this link as {\em long}
and to the transverse directions as {\em short}. Consider now a tube of
radius $r = 1 / L$ around the long direction. 
Clearly, if we constrain all the points
to lie inside the tube, the area of each triangle will not exceed
$\frac{1}{2} L r = \frac{1}{2}$. As the action is the sum of the areas
squared of the individual triangles, it will be bounded from above by
$\frac{1}{4} n_T$ {\em independently} of $L$.
Assuming that all the vertices are free to move independently inside the tube,
we can estimate the contribution to the partition function
coming from this configuration as~:
\begin{equation}\label{tube}    
Z_{tube}\sim 
\int \text{d} L \ L^3 \left(\text{d} L \ \text{d}r^3\right)^{n-2}
 D(L,r)^\gamma . 
\end{equation}
Here, $D (L, r)$ is the fermionic determinant for the tube--like
configuration. We have to integrate over only $n - 1$ vertex positions
because we already fixed one of them. By choosing a convenient co-ordinate
system, we can set this fixed vertex to the origin, and the vertex at
the other end of the link to $(L, 0, 0, 0)$. This naturally defines the
direction of the needle. Since the needle may point in any direction 
in four-dimensional space, integration over the position of this point
contributes a factor $dL L^3$ to the integral. Each of the $n - 2$
remaining vertices then contributes one integration over $L$ and three
integrations over short components, resulting finally in \eqref{tube}.

The determinant $D (L, r)$ is a uniform polynomial of degree $2 (n - 1)$, 
so one would expect at first glance that in the large $L$ limit the 
leading contribution should be $L^{2 (n - 1)}$. However, as we will show
in the appendix, because of additional symmetries of the matrix all
terms containing only long directions cancel, and we get
$D (L, r) \sim (r^2 L^{n - 3})^2$ for even $n$ and
$D (L, r) \sim (r^4 L^{n - 5})^2$ for odd $n$.
Inserting this into \eqref{tube} and integrating over all but one long
component -- which corresponds to substituting $\int dL \rightarrow L$
and $\int dr \rightarrow 1 / L$ -- we eventually obtain~:
\begin{equation}        
Z_{tube}\sim d L \ L^{-\alpha} \, , \ \rm{where} \ 
\left\{ \begin{array}{ll}
\alpha = 2n (1 - \gamma) + 10 \gamma - 7 & \rm{for \ even} \ n
\\ & \\
\alpha = 2 n (1 - \gamma) + 18 \gamma - 7 & \rm{for \ odd} \ n 
\end{array} \right.
\label{pL}
\end{equation}
Because the configuration is essentially one--dimensional and the length
of the tube is given by $L$, we expect the gyration radius to be
proportional to this, $R \sim L$. Thus, for large $R$, we expect a
similarly power--like tail in the gyration radius distribution,
\begin{equation}        
\text{d}R \ p(R)\sim \text{d}R \ R^{-\alpha} 
\end{equation}
{\em if} the needle--like configurations dominate in the partition
function. 

For the purely bosonic case $\gamma = 0$, $\alpha = 2n - 7$ grows with
the size of the system and the tail becomes less singular. This suppresses
needles. In fact, as we already discussed, in this case the spiky
configurations win entropically over the needles, and it is they who are
responsible for the singularities of the partition function for bosonic
surfaces.

For $\gamma = 1$, the formula gives $\alpha = 3$ for configurations with
an even number of vertices and $\alpha = 11$ for those with an odd number.
These exponents do not grow with $n$.

For $\gamma > 1$, the exponent $\alpha$ decreases with $n$ and very
quickly becomes smaller than one. The tail $R^{-\alpha}$ of $p(R)$ becomes
non-integrable, signalling that the partition does not exist in this case
except for a very few small systems that have $\alpha$ larger than one,
such as the tetrahedron, for which the formula predicts
$\alpha = 1 + 2\gamma$. (This result for the tetrahedron can actually be
derived in a straightforward analytic calculation, as shown in the appendix.)

To test these predictions, we measured the exponent $\alpha$ numerically
for different values of $n$ and $\gamma$. Numerical simulations for problems
with  a power--like distribution are difficult because in this case
one has  rare but important events. In our case, the distribution 
$p(R) \sim R^{-\alpha}$ means that we expect rare,
long excursions in the configuration space which produce large $R$'s. 
Technically, it is difficult to explore this part of the phase space,
because the bulk of the probability distribution is concentrated around
small $R$ where the program spends almost the whole time. To prevent this,
we introduced a lower limit $R_{min}$ on $R$, as well as an upper limit
$R_{max}$ to prevent the program from doing too long excursions to large
values of $R$. Otherwise, the program would not have a well defined
autocorrelation time. Experimentally we found it convenient to set the
limits $R_{min}=4$ and $R_{max}=10$ when we expected a negative $\alpha$,
and only an upper limit $R_{max}=7$ when we expected the distribution
$p(R)$ to grow with $R$. In figure \ref{fig:hist}, we show two examples
of the numerically obtained distribution $p(R)$~: for $n=16$ and $\gamma=1$,
where according to the formula (\ref{pL}) we expect $\alpha=3$, and for $n=8$
and $\gamma=2$, where we expect $\alpha=-3$. (Note that in the latter case,
the partition function would be ill-defined without the upper limit imposed
on $R$. We consider it here only for the purpose of testing our formula.)
\begin{figure}
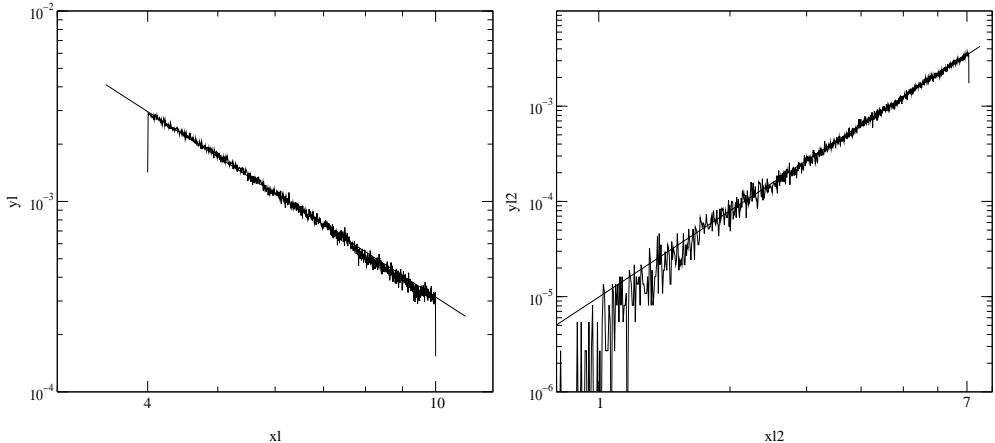

\begin{center}
\includegraphics[clip,width=6.5cm]{hist_1_16.eps}
\includegraphics[clip,width=6.5cm]{hist_2_08.eps}
\caption{\label{fig:hist}The distribution $p(R)$ for $n=16$ and $\gamma=1$ (left) and $n=8$ and $\gamma=2$ (right).}
\end{center}
\end{figure}
The numerical results for $\alpha$ are summarized in table~\ref{tab:alpha}. 
\begin{table}
\begin{center}
\begin{tabular}{|c|c|c|c|}
\hline
$\gamma$ & $n$ & $\alpha_{num}$ & $\alpha_{the}$ \\
\hline
1 & 12 &  2.39(4)  &  3 \\
1 & 16 &  2.41(3)  &  3 \\
\hline
1 & 7  &  5.20(1)  & 11 \\
1 & 11 &  3.25(1)  & 11 \\
\hline
2 & 6  &  0.92(2)  &  1 \\
2 & 8  & -3.02(5)  & -3 \\
2 & 10 & -7.12(12) & -7 \\
\hline
\end{tabular}
\end{center}
\caption{\label{tab:alpha}Values of the exponent $\alpha$ for various combinations of $n$ and $\gamma$.}
\end{table}

We can divide these results into three different cases. For $\gamma = 2$,
where needles are expected to not just create a tail in the distribution
but instead dominate it entirely, the agreement is wonderful, and there is
no question that our formula is correct. For $\gamma = 1$ and configurations
with an even number of vertices, the results are somewhat off, but still in
the ballpark of what we expect. One possibility is that this discrepancy is
due to finite size effects, and would diminish if we were to go to larger
systems; another is that our theoretical argument is simply too rough and
works only as a first approximation. In any case, we know from the analysis
of sample configurations, as presented in figure \ref{needle}, that we do
have needles in this case. For $\gamma = 1$ and an odd number of vertices,
the difference is huge but not surprising, since the analysis shows that
these configurations are not needles, but look rather like small clumps with
a single long spike growing out of them. It would seem that these things are
suppressed by a power that is smaller than $-3$ but still larger than $-11$,
so that they can dominate over needles in odd-$n$ cases but not in even-$n$
ones.

The estimate of the large $R$ behaviour of the gyration radius distribution
for needle-like configurations can easily be extended to the $d = 6$ and
$d = 10$ cases. Since we saw in the four--dimensional case that needles are
dominant only for even $n$, we restrict ourselves to these systems. If we
repeat the counting of long and short degrees of freedom as in \eqref{tube},
we get with $r \sim 1 / L$~:
\begin{equation}
Z_{tube}\sim 
\int \text{d} L \ L^{d-1} 
\left(\text{d} L \text{d}r^{(d-1)}\right)^{n-2} D(L,r)^\gamma \sim
\int \text{d} L \ L^{1 + (3-n)(d-2)} D(L,1/L)^\gamma
\label{Dtube}    
\end{equation}
where $D (L, r)$ is now the determinant 
or the Pfaffian of the Weyl fermion matrix in $d=6$ and $d = 10$,
respectively. The function $D (L, r)$ is a uniform polynomial of the
bosonic components of order \mbox{$(d - 2) (n - 1)$} for $d = 4, 6, 10$.
Because of zero modes occurring in the minimal blocks of the fermionic
matrix (as discussed in detail for $d = 4$ in the appendix), the terms
with maximal power of $L$ cancel and the leading terms for large $L$ 
are $\left( r^2 L^{n - 3} \right)^{d - 2}$. Thus, for the tube we get
$D (L, 1 / L) \sim L^{(n - 5) (d - 2)}$. Inserting this into
\eqref{Dtube} leaves us with
\begin{equation}
Z^d_{tube}\sim 
\int \text{d} L \ L^{1 - 2(d-2) + (\gamma-1)(n-5)(d-2)} \, .
\label{pLd}
\end{equation}
In the susy case, {\em i. e.} $\gamma = 1$, the tail of the
probability distribution $p(L) \sim L^{1- 2(d-2)}$
is independent of $n$. It is the same behaviour as the
one obtained in the matrix model from the analysis of the
spectrum of large eigenvalues \cite{ks}.

For $\gamma > 1$, the needle-like configurations lead to a
non--integrable singularity and the partition function
is divergent in this case. For $\gamma = 0$, the singularity
becomes softer when $n$ goes to infinity. We know, however,
that in this case needles do not play any important
role in the picture.

To summarize, the discussion presented above allows predictions for
the singular behaviour of the partition function in those cases when
needle--like configurations dominate in the ensemble. 

\section*{Discussion}

We have investigated geometrical properties of 
the large $N$ limit of the IKKT matrix model, 
which corresponds also to the reduced supersymmetric 
Yang-Mills theory \cite{ek}. We showed that the originally
four-dimensional theory reduces to a one-dimensional one 
and is dominated by elongated tube-like configurations.
The gyration radius distribution has a power--like tail
$p(R) \sim R^{-\alpha}$, where $\alpha$ is roughly consistent
with the value of $3$ obtained from the theoretical
power--counting for needle-like configurations.
Repeated in $d = 6, 10$ the power--counting for such 
one--dimensional configurations gives distributions
$p(R) \sim R^{-7}$ and $p(R) \sim R^{-15}$, respectively. 
Strikingly, the same behaviour is expected from the
analysis of the large $R$ singularities of the 
matrix model Yang-Mills integrals \cite{ks}.
This result has never been proven for the matrix model 
(except for $N=2$), but if it holds, in view of our geometrical
picture it would mean that the Yang--Mills integrals are also 
dominated by one--dimensional structures for large $R$.

It is clear from our considerations that 
if too many fermionic degrees of freedom 
were added to the model ($\gamma>1$),
the power $\alpha$ would decrease with $n$ 
and the partition function
would soon become divergent. On the other hand, 
if there were too few degrees
of freedom, as in the purely bosonic case, the singularity of the
needle--like configurations would become softer for larger $n$
similarly as in the purely bosonic matrix model \cite{hnt}.
We know, however, that in this case the surface model has
a different type of singularity, corresponding to the local creation of
spikes, caused by a sort of local flat direction in the action. The effect
introduces a stronger singularity that, again, makes the partition function
ill-defined. In summary, only the proper susy balance between fermionic and
bosonic degrees of freedom guarantees the existence of the surface theory. 

The results presented in this work are based on numerical simulations
of rather small systems. It would be very interesting to perform
a systematic analysis of the finite size effects.
This would, however, require a considerable effort to improve
the fermionic algorithm.

\section*{Acknowledgments}

The work was partially supported by the DFG PE 340/9-1 grant,
the EC IHP network {\it HPRN-CT-1999-000161} and the 
KBN grants 2P03B00814 and 2P03B14917. P.B. acknowledges
financial support from the Alexander von Humboldt Foundation 
during the first stage of this work.

\section*{Appendix}

In the first part of the appendix, we will shortly repeat 
calculations of the partition function $z_4(\gamma)$ 
for the tetrahedron \cite{oy}, and extend them to
calculate the probability distribution of the link length.

Introduce a co-ordinate system
in the target space such that the origin coincides 
with one of these vertices, setting {\em e. g.} 
$X_4 = (0, 0, 0, 0)$. Next, choose the four axes --
called here $L, x, y,$ and $z$ -- such that one other vertex lies on the $L$
axis, one lies in the $(L, x)$ plane, and one in the $(L, x, y)$ hyperplane.
Thus, we have $X_1 = (L_1, 0, 0, 0)$, $X_2 = (L_2, x_2, 0, 0)$,
and $X_3 = (L_3, x_3, y_3, 0)$.

The determinant $D(X)$ in this co-ordinate system becomes \cite{oy}
\begin{equation}
D(X) = L_1^2 \ x_2^2 \ y_3^2
\label{tetdet}
\end{equation}
Inserting this into \eqref{pf} and integrating out all co-ordinates except
$L_1$ and $x_2$ leads to
\begin{equation}
z_4 (\gamma) = \int_0^\infty d L_1 \int_0^\infty d x_2
\left( \frac{L_1^4 x_2^3}{\sqrt{\frac{3}{4} L_1^2 + x_2^2}^3} \right)^\gamma
e^{-\frac{4}{3} L_1^2 x_2^2}
\label{loso}
\end{equation}
For large $L_1$, $x_2$ is of order $1/L_1$, fixed by
the combination $L_1^2 x_2^2$ in the exponent. 
The integration over $x_2$ therefore effectively corresponds to setting
$x_2 \sim L_1^{-1}$ (including in the measure $d x_2$).
Thus, for large $L_1$, the resulting integral 
is dominated by the contribution
\begin{equation}
z_4 (\gamma) = \int_0^\infty d L_1 \ p(L_1)
\sim \int_0^\infty d L_1 \ L_1^{-1-2\gamma}
\end{equation}
The component $L_1$ of the vertex $X_1$ is the length $L$ of the link
between vertices $1$ and $4$. Since there is nothing to
distinguish between different links, 
 $p(L) \sim L^{-1-2\gamma}$ 
can be viewed as the probability
distribution of the link length for large $L$. 

As can be seen from \eqref{tetdet}, for the tetrahedron the determinant
contains short components to the fourth power, namely $x_2^2 y_3^2$. 
We will show that this feature of the determinant is not
particular to the tetrahedron, but appears in any needle-like configuration
with an even number of vertices, whereas for any needle with an odd number
of vertices we find eight short components. The reason for this lies in the
Dirac structure of the fermionic matrix.

To see this, let us define a matrix
\begin{equation} 
A' = A'_0 - i \epsilon A'_1 = 
\left( \begin{array}{cc} {f'}^0 & 0     \\ 
                         0      & {f'}^0  \end{array} \right)
- i \epsilon 
\left( \begin{array}{cc} {f'}^3            & {f'}^1 + i {f'}^2 \\ 
                         {f'}^1 - i {f'}^2 & - {f'}^3          \end{array}
\right)
\label{A}
\end{equation}
which for $\epsilon = 1$ is equal to the fermionic matrix, $A' = -a'$.
The matrices ${f'}^\mu$ are antisymmetric $(n-1) \times (n-1)$ matrices,
${f'}^\mu = -({f'}^\mu)^T$. By construction, for needle--like configurations
all elements of ${f'}^0$ are linear in the long components $L$, while those
of the ${f'}^k$ are linear in the short ones, $r_k$. Our aim is to prove
that the leading term of the determinant
$\det A'$ is proportional to at least $\epsilon^4$ for even configurations
and $\epsilon^8$ for odd, which is equivalent to saying that the leading
term contains at least four/eight small components, which in turn gives
$D (L, r) \sim r^4 L^{2n-6}$ and $D (L, r) \sim r^8 L^{2n-10}$ for even
and odd configurations, respectively. We will use first order perturbation
theory to show this.

First of all, note that ${f'}^0$ has at least one zero eigenvalue, because for
strictly one-dimensional configurations with $X_i = (X_i^0, 0, 0,  0)$ the
action \eqref{zz} has an additional zero mode coming from invariance under
a change $\psi_i \rightarrow \psi_i + \epsilon X_i^0$. (This is
a discrete remnant of a more general invariance
$\psi (\xi) \rightarrow \psi (\xi) + \epsilon g (X^0 (\xi))$ of the
continuous action, with an arbitrary function $g (X^0)$.) For the
truncated matrix ${f'}^0$, this corresponds to a zero eigenvector
\begin{equation}
w_i = X^0_i - X^0_n   \quad i = 1, \dots, n - 1
\end{equation}
Thus, we also have $\det A'_0 = (\det {f'}^0)^2 = 0$, which already excludes
terms of order $\epsilon^0$.

Consider now separately the case of even $n$. Because ${f'}^0$ has the
zero eigenvector $w$, $A'_0$ has at least two eigenvectors $u$ and $d$,
which we can write as
\begin{equation}\label{ud}
u = \left( \begin{array}{c} w \\ 0 \end{array} \right) \quad , \qquad
d = \left( \begin{array}{c} 0 \\ w \end{array} \right) \quad
\end{equation}
Under a small perturbation $-i \epsilon A'_1$ the eigenvalues 
$\lambda$ of $A'_0$ are
changed to $(\lambda - i \epsilon \Delta, \lambda + i \epsilon \Delta^*)$.
Denote the first order correction to $\lambda = 0$ by $\Delta_0$. The
determinant to this order is then proportional to
$\epsilon^2 \Delta_0 \Delta_0^* \prod' \lambda^2$, where the product runs
over nonzero eigenvalues of $f^0$. $\Delta_0$ fulfills the equation
\begin{equation}
\det \left( \begin{array}{cc} u^T A'_1 u - \Delta_0  &  u^T A'_1 d \\
                              d^T A'_1 u             &  d^T A'_1 d - \Delta_0 
\end{array} \right) \ = \ 0
\end{equation}
which gives $\Delta_0 = 0$ due to the antisymmetry of the ${f'}^k$~: $w^T
{f'}^k w =0$ for any $k$.  Thus, the eigenvalue $\lambda = 0$ remains
intact to first order, which means that all terms of order
$\epsilon^2$ likewise vanish in the determinant of $A'$. Therefore, the
leading terms must be at least of order $\epsilon^4$.  One can check
that, generically, they are indeed of this order.

For odd $n$, ${f'}^0$ is an even by even antisymmetric matrix and, as such,
cannot have just one zero eigenvector but must have at least two, which
we will call $v$ and $w$. This in turn means that $A'_0$ has at least
four such eigenvectors, which we write as
\begin{equation}
u_1 = \left( \begin{array}{c} v \\ 0 \end{array} \right) \ , \
u_2 = \left( \begin{array}{c} w \\ 0 \end{array} \right) \ , \
d_1 = \left( \begin{array}{c} 0 \\ v \end{array} \right) \ , \
d_2 = \left( \begin{array}{c} 0 \\ w \end{array} \right)
\end{equation}
Repeating the perturbation analysis in $\epsilon$ for the matrix $A'$
\eqref{A}, we find for the first order correction $\Delta_0$~:
\begin{equation}
\left[ \Delta_0^2 + \sum_{i=1}^{3} \left( v^T {f'}^i w \right)^2 \right]^2 
= 0 \, .
\end{equation}
We will now argue that all three terms in the sum vanish, and therefore
$\Delta_0 = 0$. This follows from the peculiar structure of the matrices
${f'}^\mu$. As we know, their elements are linear in the components $X^\mu$,
and moreover all the matrices have the same structure. Therefore, we can
introduce a linear function ${f'}$ such that ${f'}^\mu = f(X^\mu)$. Also, the
eigenvector $w$ is likewise linear in $X^\mu$, and we can again write
$w^\mu = w(X^\mu)$. Now we have
\begin{eqnarray}
v^T {f'}^i w & = & v^T (X^0) \, {f'} (X^i) \, w (X^0) \nonumber \\
& = & v^T (X^0) \, {f'} (X^i) \,
\Big( w (X^0) + w (X^i) \Big) \nonumber \\
& = & v^T (X^0) \, \Big( {f'} (X^0) + {f'} (X^i) \Big) \,
\Big( w (X^0) + w (X^i) \Big) \nonumber \\
& = & v^T (X^0) \, {f'} (X^0 + X^i) \, w (X^0 + X^i) \nonumber \\
& = & 0
\end{eqnarray}
(In the first step, we just added $v^T {f'} (X^i) \, w (X^i) = 0$; in the
second, we used the fact that $v^T (X_0) \, {f'} (X_0) = 0$; and in the
third, we used the linearity of ${f'}$ and $w$.)

Thus, we have shown that there are still four zero eigenvalues even to
first order of $\epsilon$. As with even $n$, we checked numerically that
they are non-zero to the second order. Thus, the leading term in this
case behaves as $\epsilon^8$.

\end{document}